\documentclass[twocolumn,amsmath,amssymb,aps,pre,reprint,longbibliography]{revtex4-1}
\usepackage{graphicx}
\begin{document}

\title{
Clogging, Dynamics and Reentrant Fluid For Active Matter on Periodic Substrates 
} 
\author{
C. Reichhardt and C. J. O. Reichhardt 
} 
\affiliation{
Theoretical Division and Center for Nonlinear Studies,
Los Alamos National Laboratory, Los Alamos, New Mexico 87545, USA\\ 
}

\date{\today}
\begin{abstract}
We examine the collective states of run-and-tumble active matter disks driven over a periodic obstacle array. When the drive is applied along a symmetry direction of the array, we find a clog-free uniform liquid state for low activity, while at higher activity, the density becomes increasingly heterogeneous and an active clogged state emerges in which the mobility is strongly reduced. For driving along non-symmetry or incommensurate directions, there are two different clogging behaviors consisting of a drive dependent clogged state in the low activity thermal limit and a drive independent clogged state at high activity. These regimes are separated by a uniform flowing liquid at intermediate activity. There is a critical activity level above which the thermal clogged state does not occur, as well as an optimal activity level that maximizes the disk mobility. Thermal clogged states are dependent on the driving direction while active clogged states are not. In the low activity regime, diluting the obstacles produces a monotonic increase in the mobility; however, for large activities, the mobility is more robust against obstacle dilution. We also examine the velocity-force curves for driving along non-symmetry directions, and find that they are linear when the activity is low or intermediate, but become nonlinear at high activity and show behavior similar to that found for the plastic depinning of solids. At higher drives the active clustering is lost. For low activity we also find a reentrant fluid phase, where the system transitions from a high mobility fluid at low drives to a clogged state at higher drives and then back into another fluid phase at very high drives. We map the regions in which the thermally clogged, partially clogged, active uniform fluid, clustered fluid, active clogged, and directionally locked states occur as a function of disk density, drift force, and activity.  
\end{abstract}
\pacs{75.70.Kw,75.70.Ak,75.85.+t,75.25.-j}
\maketitle

Active matter describes systems that have some form of self-propulsion,
such as biological entities, artificial swimmers, and robots.
This area is attracting
increasing attention due to the experimental realization
of new types of active matter 
including individual or collective elements in
uniform or complex environments \cite{Marchetti13,Bechinger16}.
One class
of active matter is composed of
assemblies of self-driven particles such as disks or elongated objects,
in which the activity can arise 
from driven diffusion, run-and-tumble dynamics, chiral motion,
or other forms of mobility \cite{Marchetti13,Bechinger16}. 
In disorder-free environments,
one of the most striking features of such particle-based
active matter is the appearance
of motility-induced phase separation
\cite{Fily12,Redner13,Palacci13,Buttinoni13,Cates15,Reichhardt14a}, 
which occurs when the activity is large enough
to induce a transition 
from a uniform fluid to a coexistence of 
dense or solid regions with a low density active gas.
The dense regions form even when
all of the pairwise interactions between the particles are repulsive. 

When active matter particles are coupled to some form of complex environment such as walls, 
obstacles, or pinning sites, additional behaviors can arise.
These include
the induced locomotion of passive objects as well as
a wetting phenomenon in which active particles can
align with a barrier or accumulate in corners
\cite{Yang14,Mallory14,Takatori14,Ray14,Ni15,Reichhardt17a,Solon15,Speck20}.
If the active particles interact with a random array 
of obstacles, pins, or other types of quenched disorder,
the clustering or flocking transition can be destroyed, and
under
application of a drive
can be replaced
by a transition from isotropic to anisotropic cluster phases
\cite{Chepizhko13,Morin17,Sandor17b,Yllanes17,Sandor17a,Chardac20}. 
It is also possible for different types of trapping effects to appear
\cite{Bhattacharjee19a,Shi20}, as well
as transport or hopping rates that are non-monotonic as a function of
activity \cite{Reichhardt14,Morin17a,Bertrand18}.
The non-monotonicity arises because
at low activity the system is easily pinned or clogged,
at intermediate activity it acts like a freely flowing fluid,
and at high activity it
undergoes motility-induced phase separation 
accompanied by
active clogging in which
the motion is strongly reduced and occurs
only intermittently or in avalanches
\cite{Reichhardt14,Reichhardt18c,Reichhardt18a}.

It is also possible to place active matter particles
on a periodic array of obstacles or a periodic pinning
array \cite{Volpe11,Lozano16,BrunCosmeBruny20,Reichhardt20}.
In this case,
there is a well defined
average distance between the obstacles as
well as clear preferred symmetry directions for motion.
For non-active particles driven over periodic arrays, strong
directional locking effects
occur when the direction of the drive is changed with respect to the symmetry of 
the underlying substrate
\cite{Reichhardt20,Reichhardt99,Korda02,Gopinathan04,Balvin09,Koplik10}.
The particles
preferentially flow along the symmetry directions of the substrate
even when the external drive is not aligned with one of those directions.
As a result,
a series of
locking steps
appear in the velocity versus
driving angle curves,
as has been studied for
colloidal
\cite{Korda02,Gopinathan04,Balvin09,Reichhardt11,Bohlein12a,Stoop20,Cao19}
and other driven particle systems
\cite{Reichhardt20,Silhanek03,Vizarim20,Feilhauer20}
coupled to a periodic substrate. 
The locking can be used
as a method for
steering the motion of particles along
certain directions
or for performing
particle separation
in which one species locks to one symmetry direction and 
another species locks to another direction
\cite{MacDonald03,Huang04,Risbud14,Roichman07a,Long08}.
Directional locking is the most pronounced in the single particle limit,
where
the existence of multiple locking directions
produces
a devil's staircase
structure in the velocity versus angle response;
however, 
when collective effects become important,
different types of pattern formation can arise in which
ordered flowing solids are stable for
motion along symmetry locking directions 
but liquid or disordered states appear for flow
along incommensurate angles
\cite{Reichhardt20,Reichhardt11,Bohlein12a,Stoop20,Cao19,Reichhardt20a}.

There have been several experimental \cite{Volpe11}
and numerical studies
\cite{Reichhardt20,Zeitz17,Pattanayak19,AlonsoMatilla19,Schakenraad20,Ribeiro20,Reichhardt21} 
of active matter systems interacting with a periodic substrate.
At large activity in the single particle
limit,
the active disk motion 
is directionally locked even in the absence of
external driving, and the prominence of this locking
effect increases with increasing run times \cite{Reichhardt20}.
For a driven or directed active matter system,
the activity can 
enhance
the directional locking effect \cite{Volpe11,BrunCosmeBruny20,Reichhardt20}. 

In this work we expand upon our previous studies
of individual active matter disks
interacting with periodic obstacle arrays \cite{Reichhardt20},
and consider
the strongly collective regime 
in which clogging effects appear at both
very low and very high activity.
We show that different types of clogging
occur depending on the direction of
drive with respect to the symmetry directions of the substrate.
For driving along
a major symmetry direction such as $\theta = 0^\circ$,
the system forms
a uniform liquid state at low activity and
the mobility monotonically decreases with increasing activity until
an activity-induced cluster state emerges in which the mobility is
strongly reduced.
For driving along an incommensurate angle,
there can be
two distinct clogging phases.
The first, which we call thermal clogging, is 
a low activity clogged state similar
to the clogging studied
in previous work
for zero activity
\cite{Nguyen17,Peter18} and  
for the low activity regime
of active matter on random substrates
\cite{Reichhardt21a,Stoop18,Leyva20}, where the
active particles
form a heterogeneous system-spanning cluster.
The second is a high activity state
called active clogging which is associated with motility
induced phase separation.
There is a critical activity level
above which the thermally clogged state is lost. 
Under
a finite drive,
both the thermal and active clogging
depend on the active disk density.  
We examine the role of the driving force
and find that active clogging persists
down to the smallest drives since the
activity itself induces the clustering effect.
At higher drives, the active clogging is diminished since
the drive interferes with cluster formation.
In the low activity limit, we observe a reentrant fluid phase
in which the system
is a uniform fluid at low drives,
transitions into a drive-induced clogged state,
and breaks apart into a moving fluid or partially clustered state at
high drives.
These transitions are associated with
a finite depinning threshold and negative differential mobility. 
For the active clogged state,
there is no depinning threshold but
the velocity-force curves
develop nonlinearity
at low drives
similar to that found at plastic depinning in
a non-thermal system, while at intermediate activity the
velocity-force curves are linear. 
At high driving
along incommensurate angles,
for both low and high activity
a directionally locked state
appears in which
the clustering
is lost and the disks flow in strictly
one-dimensional (1D) channels.     
In this work we focus on a fixed obstacle size and density.
The collective effects that appear
for changing obstacle size were studied 
in earlier work \cite{Reichhardt21}, where
it was shown that commensuration effects arise
when the spacing of the obstacle lattice matches with
the active disk size,
permitting the formation of large scale crystalline structures. 

\section{Simulation}
We consider a two-dimensional (2D)
system of size $L \times L$
with periodic boundary conditions in the $x$ and $y$ directions.
The sample contains a square array of $N_{\rm obs}$
obstacles of radius $R_{\rm obs}$
and lattice spacing $a$, giving an obstacle density of
$\phi_{\rm obs}=N_{\rm obs}\pi R^2_{\rm obs}/L^2$.
There are also $N_{a}$ active disks
with a radius of $R_{a}$.
The density of the active disks 
is   
$\phi_{a} = N_{a}\pi R^2_{a}/L^2$,
and the total system density is $\phi_{\rm tot} = \phi_{a} + \phi_{\rm obs}$. 
After initialization we apply a drift force
$F_{D}$ along either the $x$ direction at $\theta=0^\circ$, which is
a symmetry direction of the obstacle array, or
along
$\theta = 35^\circ$, which is a non-symmetry direction.
In our previous work with this system in the zero activity limit 
at higher $\phi_{\rm tot}$,
we found that the disks are susceptible
to clogging when $\theta = 35^\circ$ but generally 
flow when $\theta = 0^\circ$
\cite{Nguyen17}.

The equation of motion for an active disk $i$ is given by
\begin{equation} 
\alpha_d {\bf v}_{i}  =
{\bf F}^{dd}_{i} + {\bf F}^{m}_{i} + {\bf F}^{obs}_{i} + {\bf F}^{D} \ .
\end{equation}
Here ${\bf v}_{i} = {d {\bf r}_{i}}/{dt}$ is the disk velocity, 
${\bf r}_{i}$ is the disk position, and the damping constant $\alpha_d = 1.0$.
The active disks are modeled as
run-and-tumble particles
with strong harmonic repulsive interactions
of the form 
${\bf F}^{dd}_{i} = \sum_{i\neq j}^{N_a}k(2R_{a} - |{\bf r}_{ij}|)\Theta(2R_{a} - |{\bf r}_{ij}|) {\hat {\bf r}_{ij}}$,
where $\Theta$ is the Heaviside step function,
${\bf r}_{ij} = {\bf r}_{i} - {\bf r}_{j}$, and
$\hat {\bf r}_{ij}  = {\bf r}_{ij}/|{\bf r}_{ij}|$. We
fix the spring constant to $k = 100$,
which is large enough that the  
system behaves close to the hard disk limit
over the range of forces we consider.
The obstacles are also modeled as disks with
the same form of harmonic repulsion
but with a different radius $R_{\rm obs}$. 
In this work we fix
$R_{\rm obs} = 1.0$, $R_{a} = 0.45$, and the obstacle lattice constant 
$a = 3.0$, giving $\phi_{\rm obs}=0.349$.
We consider
varied system sizes between $L = 36$ and $L=72$, but unless
otherwise noted, the system size is $L=36$.
The activity of the disk is produced by
a motor force ${\bf F}^{m}_i=F_M{\hat {\bf r}}^m_i$
where $F_M=1.0$ and
${\hat {\bf r}}^m_i$ is a randomly chosen direction
which changes instantaneously every
$\tau_{l}$ time steps.
All of the active disks use the same value of $\tau_l$, but the
running of each disk is
started at different randomly chosen times in the cycle
so that
the disks do not all tumble simultaneously.
We quantify
the activity using the run length $l_{r}$ which is
the distance a single isolated active
disk would move during
the run time $\tau_{l}$.
In general, we find that when
$l_{r} < 0.1a$, the behavior is close to the thermal limit,
as studied previously
in the single particle limit  \cite{Reichhardt20}.    
A uniform drift force ${\bf F}^{D}=F_D{\bf \hat{d}}$
is applied to all the active disks,
where ${\bf \hat{d}}$ is along either $\theta=0^\circ$ or $\theta=35^\circ$.

We characterize the transport by measuring
$\langle V_{x}\rangle = N_a^{-1}\sum^{N_a}_{i= 1}{\bf v}\cdot {\bf \hat{x}}$
and $\langle V_{y}\rangle = N_a^{-1}\sum^{N_a}_{i= 1}{\bf v}\cdot {\bf \hat{y}}$.
The net velocity
is given by
$\langle V\rangle = (\langle V_x\rangle^{2} + \langle V_y\rangle^{2})^{1/2}$.
We define
the mobility of the system
as $M = \langle V\rangle/F_{D}$. 

\section{Results}

\begin{figure}
\includegraphics[width=\columnwidth]{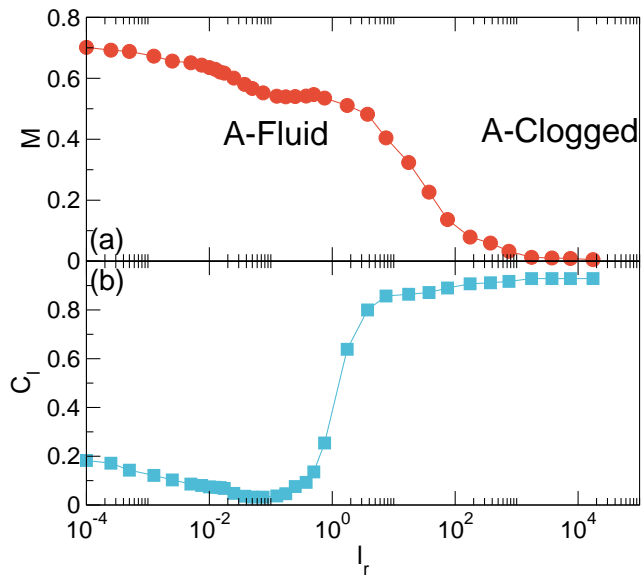}
\caption{ 
(a) Mobility $M$ vs active run length $l_{r}$ for
  a
  system with a square obstacle array
under a drift force $F_D=0.1$ applied at $\theta = 0^\circ$
at
a total density of $\phi_{\rm tot} = 0.622$.
(b) The corresponding fraction of disks $C_l$ that
are in the largest cluster
vs $l_{r}$.
$M$ is large at low $l_r$ and decreases with increasing $l_r$, exhibiting
a sharper drop that correlates
with an increase in $C_{l}$.
The active fluid (A-Fluid) and active clogged (A-Clogged) regimes are
labeled.
}
\label{fig:1}
\end{figure}

\begin{figure}
\includegraphics[width=\columnwidth]{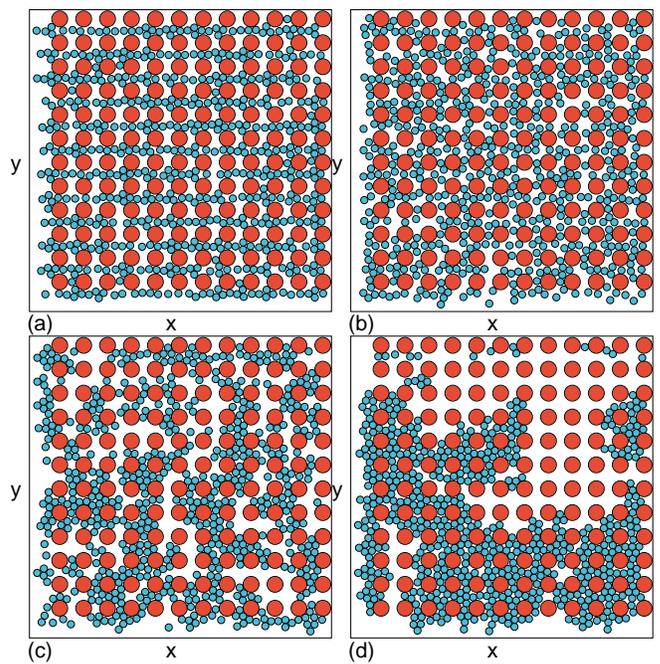}
\caption{
  Snapshots showing the active disk positions (blue) and obstacle
  locations (red)
  for the system in Fig.~\ref{fig:1} with $F_D=0.1$,
  $\phi_{\rm tot}=0.622$, and $\theta=0^\circ$.
(a) 1D channel formation at $l_{r} = 0.00025$.
(b) A uniform liquid at $l_{r} = 0.175$.
(c) At $l_{r} = 17.5$, clustering begins to occur
and the mobility $M$ is reduced. 
(d) An active clogged state at $l_{r} = 7500$, with little motion.  
}
\label{fig:2}
\end{figure}

In Fig.~\ref{fig:1}(a) we plot the mobility $M$
versus active run length $l_{r}$ for a system with a fixed
external drift force
of $F_{D} = 0.1$ applied along the $x$-direction
at $\theta=0^\circ$
for
$\phi_{\rm tot} = 0.622$.
Figure~\ref{fig:1}(b) shows the corresponding fraction of disks that 
are in the largest cluster $C_{l}$
versus $l_{r}$.
To measure $C_{l}$, we identify all disks $N_c$ which are in
contact with each other in the largest cluster, and divide this
by the total number of active disks to obtain $C_l=N_c/N_a$.
In Fig.~\ref{fig:1}(a),
$M \approx 0.7$ at low drives,
indicating a high mobility,
and gradually decreases with increasing $l_{r}$.
There is a sharper
decrease in $M$
for $l_{r} > 1.0$ corresponding to
an increase in $C_{l}$,
which indicates
the onset of activity-induced clustering.
For $l_{r} > 1000$, $M$ drops to $M=0.01$,
almost two orders of magnitude smaller than its value
in the low activity regime. 

In Fig.~\ref{fig:2}(a) we
illustrate the positions of the disks and obstacles in the
system from Fig.~\ref{fig:1} for the high mobility
state with $l_{r} = 0.00025$,
where the disks move in 1D channels between the obstacles.
There is some buckling
of the channel structure,
resulting in disk-obstacle collisions which reduce the mobility to
$M < 1.0$.
As $l_{r}$ increases, the disk configurations become more disordered
and disk-obstacle collisions become more frequent.
This produces
a fluctuating uniform liquid state in which
the 1D channeling is lost and the mobility has decreased moderately, as
shown in Fig.~\ref{fig:2}(b) at $l_{r} = 0.175$.
At $l_r=17.5$ in Fig.~\ref{fig:2}(c),
an active clustering effect is beginning to occur
which causes
a reduction in the mobility to
$M = 0.35$,
while at $l_r=7500$ in
Fig.~\ref{fig:2}(d),
the system forms an almost completely frozen cluster
in what we term an active clogged state,
with a mobility close to zero and
only occasional small collective rearrangements.
As $l_{r}$ increases,
the mobility decreases further,
until in the infinite running time or ballistic limit, $M = 0$. 

The results in Fig.~\ref{fig:1} can be compared
to the behavior 
of active matter on random substrates.
In Ref.~\cite{Reichhardt14},
the drift velocity or mobility in a random
obstacle array
is generally
low at small $l_{r}$, reaches a maximum at intermediate $l_{r}$,
and then decreases again for larger $l_{r}$. 
Similarly, in Ref.~\cite{Reichhardt18c}, 
the system is in a clogged state at low $l_{r}$,
forms a liquid state for intermediate run lengths,
and enters an active clogged state
with nearly zero mobility at high drives.
In contrast, Fig.~\ref{fig:1} shows that 
there is no clogged or low mobility state
at small $l_{r}$ in an ordered obstacle array.
Here the disks follow
easy flow 1D channels 
between the obstacles
for driving along $\theta=0^\circ$ at low $\phi_{\rm tot}$, as illustrated
in Fig.~\ref{fig:2}(a), and
the mobility $M\approx 1$ since disk-obstacle collisions are
minimized.
At larger
$l_{r}$,
an active clogged state appears regardless of
whether the obstacle lattice geometry is periodic or random.

\begin{figure}
\includegraphics[width=\columnwidth]{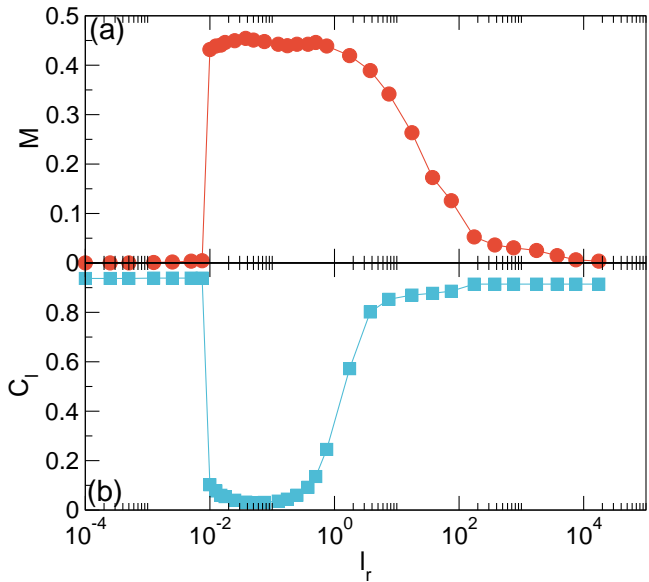}
\caption{
  (a) $M$ vs
  $l_{r}$ 
  for the system in
  Fig.~\ref{fig:1} with $F_D=0.1$
  and $\phi_{\rm tot}=0.622$
  but where the drive is applied along
  $\theta = 35^\circ$.
  (b) The corresponding $C_{l}$ vs $l_{r}$.
  Here there are two distinct clogged states:
  a thermally clogged state at low $l_{r}$,
  shown in Fig.~\ref{fig:4}(a), and
an active clogged state at large $l_r$, shown in Fig.~\ref{fig:4}(d).   
}
\label{fig:3}
\end{figure}

\begin{figure}
\includegraphics[width=\columnwidth]{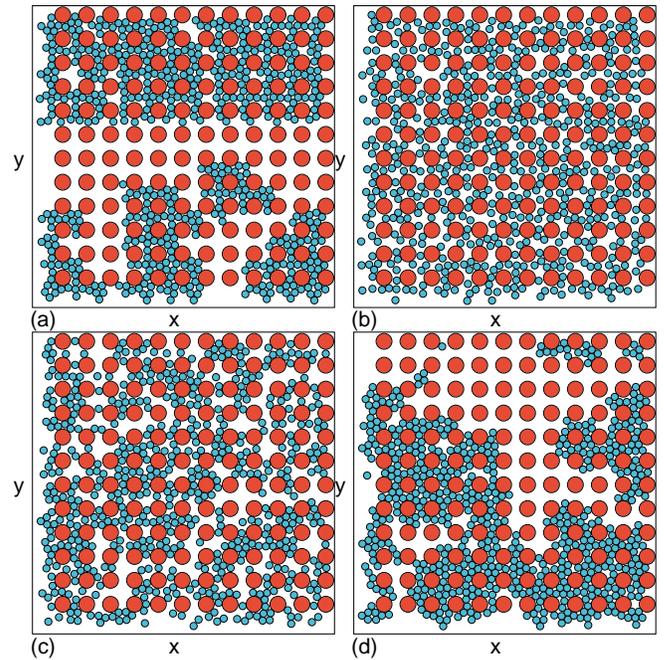}
\caption{
  Snapshots showing the active disk positions (blue) and obstacle locations
  (red) 
  for the system in Fig.~\ref{fig:3} with
  $F_D=0.1$,
  $\phi_{\rm tot}=0.622$, and
  $\theta = 35^\circ$.
(a) A thermally clogged state at $l_{r} = 0.00125$.
(b) A uniform liquid at $l_{r} = 0.175$.
(c) At $l_{r} = 3.75$, the mobility begins to drop
as clustering starts to occur.  
(d) An active clogged state at
$l_{r} = 7500$.
}
\label{fig:4}
\end{figure}

In Fig.~\ref{fig:3}
we plot $M$ and $C_{l}$ versus $l_{r}$
for the same system as in Fig.~\ref{fig:1} but with
the drive applied along
$\theta = 35^{\circ}$.
For $l_{r} < 0.01$, $M \approx 0$
and $C_{l} \approx 0.92.$
Here
we find
a non-active or thermally clogged 
state of the type shown in Fig.~\ref{fig:4}(a) at $l_{r} = 0.00125$.
The clogging arises from the formation
of bottlenecks and arches that build up over time and block the flow.
This state is similar to the $T = 0$ clogging states 
studied previously for passive particles in periodic arrays
\cite{Nguyen17,Reichhardt21a,Stoop18},
random arrays \cite{Leyva20,Reichhardt14b},
and bottleneck systems \cite{Zuriguel15,Dressaire17},
where the clogged state is associated with strongly heterogeneous
particle density.
We note that for smaller $F_{D}$ at these small values of $l_r$,
the activity is still large enough that
the system forms a liquid rather than a clogged
state,
as we describe in Section V.

For $0.01 < l_{r} < 1.0$  in Fig.~\ref{fig:3}(a,b),
a uniform liquid state appears with
$C_l<0.2$ and
a mobility of
$M \approx 0.55$,
as illustrated
in Fig.~\ref{fig:4}(b) for $l_{r} = 0.175$. There is a
drop in the mobility at
higher $l_{r}$ 
corresponding to the formation of an activity-induced cluster state
of the type shown
in Fig.~\ref{fig:4}(c) at $l_{r} = 3.75$.
For $l_r>100$, the system enters
an actively clogged state
with little motion,
as shown in Fig.~\ref{fig:4}(d) at $l_{r} = 7500$.

The  behavior of $M$ versus $l_{r}$ in Fig.~\ref{fig:3}
has differences
from previous work on
active matter driven through
random arrays at low activity \cite{Reichhardt18c}. 
The transition from the passive clustered
state to the active fluid state is much sharper in the periodic array. 
This is likely a result of the well-defined spacing between obstacles
in the periodic array. 
In general, the onset of active clustering 
occurs when the run length becomes
larger than the distance between the obstacles.
For smaller $l_{r}$,
the formation of the thermally clogged state
is sensitive to
the direction $\theta$ of the drive with
respect to the substrate symmetry, while the
active clogged state at  
large $l_{r}$ 
is independent of $\theta$.

\begin{figure}
\includegraphics[width=\columnwidth]{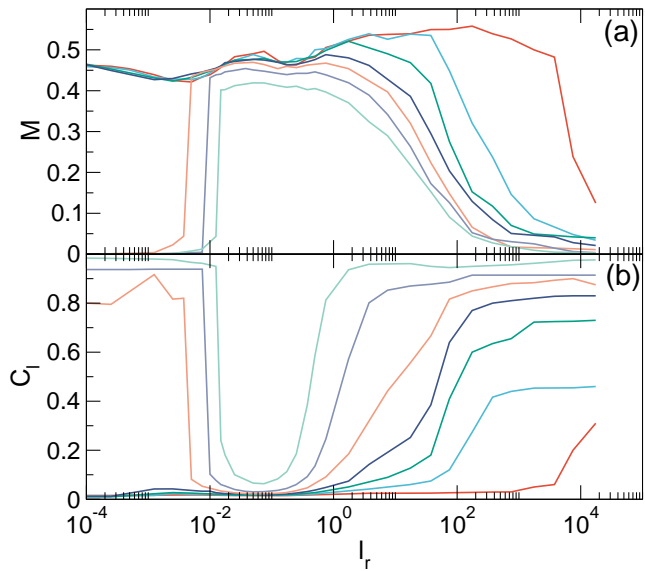}
\caption{
  (a) $M$ vs $l_{r}$ for the system in
  Fig.~\ref{fig:3} with $F_D=0.1$
  and $\theta=35^\circ$
  for varied total density
  $\phi_{\rm tot}=0.6708$ (light green),
  0.622 (light purple),
  0.573
  (orange),
  0.5238 (dark blue),
  0.4743
  (dark green),
  0.4264
  (light blue),
  and  $0.3655$ (dark red).
(b) The corresponding $C_{l}$ vs $l_{r}$. }
\label{fig:5}
\end{figure}

Since driving along different angles can produce
both thermal and active clogging, we next consider 
the evolution of the different phases 
as a function of
the active disk density $\phi_a$ for fixed $F_{D}$. 
In Fig.~\ref{fig:5}(a) we plot $M$ versus $l_{r}$ for the system in
Fig.~\ref{fig:3} with
$\theta=35^\circ$ at 
$\phi_{\rm tot}=0.6708$, 0.622, 0.573, 0.5238, 0.4743, 0.4264, and  $0.3655$, 
while in Fig.~\ref{fig:5}(b) we plot
the corresponding $C_{l}$ versus $l_{r}$.
For $\phi_{\rm tot} >  0.5238$,
a thermally clogged state appears
at small $l_{r}$,
while the critical $l_{r}$ at which the system transitions into
the active fluid phase
decreases with
decreasing $\phi_{\rm tot}$.
When $\phi_{\rm tot} \leq 0.5238$,
the mobility is almost independent of $l_{r}$ for small  
and intermediate values of
$l_r$.
For all
values of $\phi_{\rm tot}$,
there is
a transition at larger $l_r$ to an active clustered state
which is associated with a simultaneous
drop in $M$ and increase in $C_{l}$.
The maximum value of 
$C_{l}$ in the active clustered phase
gradually drops
with decreasing $\phi_{\rm tot}$ as the clusters
become more spatially separated.
The crossover
to the active clogged state
shifts to larger values of $l_{r}$ as $\phi_{\rm tot}$ decreases.

\begin{figure}
\includegraphics[width=\columnwidth]{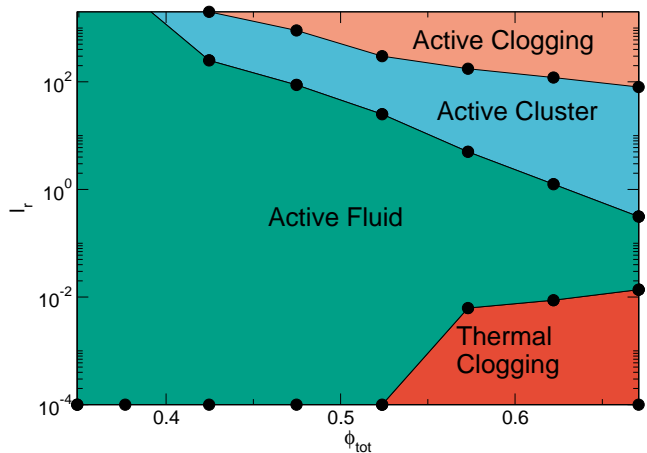}
\caption{
Phase diagram as a function of $l_r$ vs $\phi_{\rm tot}$
for the system in Figs.~\ref{fig:3} and \ref{fig:5}
with $F_D=0.1$
and $\theta=35^\circ$.
Green: the active fluid with uniform density, low $C_{l}$, and high $M$.
Red: the thermally clogged phase
with high $C_{l}$ and $M$ close to zero.
Blue: the clustered fluid phase with high $C_{l}$ and intermediate $M$. 
Orange: the active clogging phase characterized by
low mobility and high $C_{l}$. 
}
\label{fig:6}
\end{figure}

In Fig.~\ref{fig:6} we show a phase diagram as a function
of $l_r$ versus $\phi_{\rm tot}$ for the system from Figs.~\ref{fig:3} and
\ref{fig:5} with
$\theta=35^\circ$.
In the active fluid,
the mobility is high but no clustering occurs. 
For the thermally clogged state, the mobility is zero and strong clustering
occurs.
The thermally clogged state is present only for incommensurate driving
angles such as $\theta=35^\circ$, but the active clogged state is always
present for sufficiently large $l_r$ and $\phi_{\rm tot}$.
We use the value of the mobility to
draw a distinction between the active cluster state and
the active clogged state.
In the active cluster state,
$C_{l}$ is high and $M$ is larger than 15\% of its maximum value,
while in the active clogged state, $C_{l}$
is also high but $M$ is less than 15\% of its maximum value.
Other definitions for the boundary of the active clogged state
could be used that involve lower values of $M$.  
The transition from the active fluid to 
the active cluster state is best described as a crossover,
as shown in Fig.~\ref{fig:5}(a).
A similar phase
diagram can be constructed for driving at
$\theta = 0^\circ$ (not shown).
Here the thermal clogging phase is absent but 
it would be possible to define additional 1D and quasi-2D channeling
phases
at low $l_{r}$ which would be distinct from the
2D uniform active fluid state.
For large enough $\phi_{\rm tot}$
at $\theta = 0^\circ$,
the active fluid could cross over into
a uniform jammed state,
but these extremely dense states become very computationally expensive,
so we limit our study to
densities at which only heterogeneous clogging occurs.
Similarly,
at values of
$\phi_{\rm tot}$ larger than what we consider
for the $\theta=35^\circ$ system in
Fig.~\ref{fig:6}, a crossover could occur
from the thermally clogged phase to a uniform jammed phase
near $\phi_{\rm tot} = 0.9$ \cite{Reichhardt14b}.

\begin{figure}
\includegraphics[width=\columnwidth]{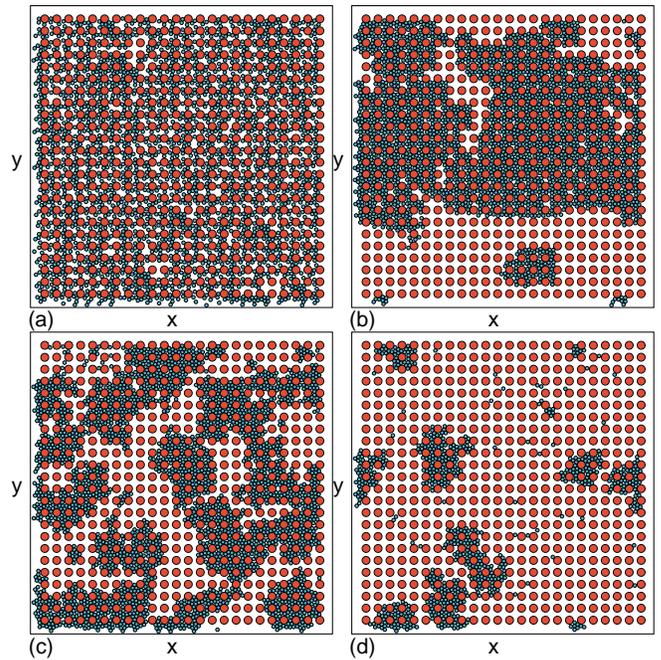}
\caption{
  Snapshots showing the active disk positions (blue) and obstacle locations
  (red)
  for driving at $\theta = 35^\circ$ for the system in
  Fig.~\ref{fig:3} with $F_D=0.1$
  in a larger sample with $L=72$.
  (a) A uniform state for $\phi_{\rm tot}=0.622$
  just after the initialization at $l_{r} = 0.00125$.
  (b) The same sample from (a) after
  the system reaches a thermally clogged state
  containing
  a large system-spanning cluster. 
  (c) The active clogged state at $l_{r} = 7500$
  and $\phi_{\rm tot}=0.622$ in which
  the clusters are not system-spanning. 
  (d) An active clogged state at $l_{r} = 7500$ and
  $\phi_{\rm tot} = 0.4265$.
}
\label{fig:7}
\end{figure}

\begin{figure}
\includegraphics[width=\columnwidth]{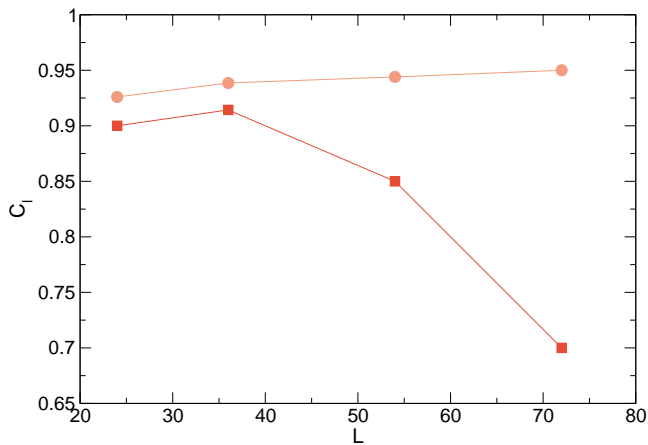}
\caption{$C_{l}$ vs system size $L$ for the system in
  Fig.~\ref{fig:3} with $F_D=0.1$,
  $\phi_{\rm tot}=0.622$,
  and $\theta=35^\circ$ at
  $l = 0.00125$ in the thermally clogged state (orange circles)
  and $l = 7500$ in the active clogged state
  (red squares).
  The thermally clogged state is dominated by a single
  system-spanning cluster while
  the active clogged state contains
  multiple smaller clusters.   
}
\label{fig:8}
\end{figure}

We have explored the robustness of the phases described above
to changes in the system size.
As shown in Fig.~\ref{fig:7}, we find similar results
for a system with the same parameters as in
Fig.~\ref{fig:3} but with a larger size of $L = 72$. 
Figure~\ref{fig:7}(a) illustrates the initial state for
$\phi_{\rm tot} = 0.622$ and $l_{r} = 0.00125$
where the active disk density is still nearly uniform.
In Fig.~\ref{fig:7}(b), the same sample has
reached a thermally clogged state
containing a large system-spanning cluster.
Figure~\ref{fig:7}(c) shows the configuration in
the active clogged state at
$\phi_{\rm tot}=0.622$ and $l_{r}= 7500$.
The morphology of the clusters in the active clogged state differs
from those in the thermally clogged state.
The active clogged clusters are
much more fragmentary and do not form
a single system-spanning cluster. 
The thermally clogged
states generally contain a
cluster that is as wide as the system,
while the active clogged state forms individual 
clusters     
that are each considerably smaller than the system width.
This suggests that the thermally clogged state
requires the total disk density
to be above a 2D percolation threshold, while
the active clogged
state does not.
In Fig.~\ref{fig:7}(d)
we show an active clogged state at
$l_r=7500$ and $\phi_{\rm tot} = 0.4265$.
At low $l_r$ for this same total density, 
the system does not form a thermally clogged state. 
In Fig.~\ref{fig:8} we plot $C_{l}$
versus $L$ for the system in
Fig.~\ref{fig:3} at $l = 0.00125$ in the thermally clogged phase
and $l = 7500$ in the active clogged phase.
For the thermally clogged phase, $C_l \lesssim 1$
independent of the system size,
indicating that the system is dominated by
a single system-spanning
cluster as shown in Fig.~\ref{fig:7}(b).
In contrast, for the active clogging phase illustrated in
Fig.~\ref{fig:7}(c),
$C_l$ drops as $L$ increases since
the size of the largest cluster does not increase with system
size; instead, the total number of clusters grows as $L$
increases, so that each cluster contains a smaller fraction of
the total number of disks.

\section{Diluted Obstacle Arrays}

\begin{figure}
\includegraphics[width=\columnwidth]{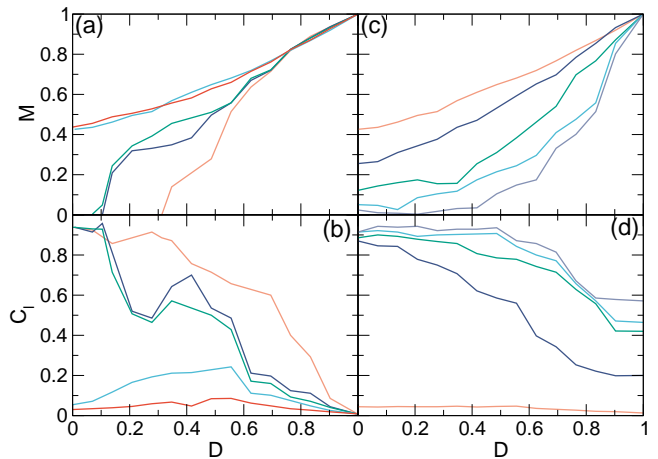}
\caption{
  (a) $M$ and (b) $C_{l}$
  vs the dilution factor $D$ for the
  system in Fig.~\ref{fig:3} with $F_D=0.1$,
  $\phi_{\rm tot}=0.622$, and $\theta=35^\circ$
at
$l_{r} = 0.00025$ (orange),
0.005 (dark blue),
0.0075 (dark green),
0.0175 (light blue), and $0.05$ (red). 
(c) $M$ and (d) $C_{l}$ for
the same system at
$l_{r} = 0.175$ (orange),
17.5 (dark blue),
75.0 (dark green),
175 (light blue), and $1750$ (light purple).
}
\label{fig:9}
\end{figure}

We next consider the effects of randomly diluting the obstacle array
by removing a fraction $D$ of the obstacles.
In Fig.~\ref{fig:9}(a,b) we plot
$M$ and $C_{l}$ versus the dilution factor
$D$ for the system in
Fig.~\ref{fig:3} with $\theta=35^\circ$,
$\phi_{\rm tot}=0.622$, and $F_D=0.1$ 
at $l_{r} = 0.00025$, 0.005, 0.0075, 0.0175, and $0.05$.
For $l_r = 0.00025$, 0.005, and $0.0075$,
the thermally clogged state with $M \approx 0$ survives at low $D$. 
The threshold for a transition to a depinned state with finite $M$
shifts to lower values of $D$
as $l_{r}$ increases.
For $l_{r} = 0.00025$,
the onset of mobility
occurs for $D \approx 0.4$ and is accompanied by
a small local peak in $C_{l}$.
As $D$ increases,
the mobility
increases monotonically,
reaching $M=1.0$ at $D = 1.0$ when there are no obstacles left
in the system.
In general, $C_{l}$ decreases
with increasing $D$ and reaches
$C_l \approx 0$
for $D = 1.0$ since the disks are no longer in contact. 
For $l_{r} = 0.005$ and $0.0075$, some clustering still persists
even when $M$ is finite due to the formation of a partially
clogged state, but
the clustering diminishes when
$D > 0.6$.  
For $l_{r} > 0.0125$, there is no 
clogged state,
$M$ is finite for all values of $D$,
and the clustering is strongly reduced.

In Fig.~\ref{fig:9}(c,d) we plot $M$ and $C_{l}$ versus $D$ for
the same system in Fig.~\ref{fig:9}(a,b) at
$l_{r} = 0.175$, 17.5, 75.0, 175, and $1750$.
Here 
the average value of $M$ decreases
with increasing $l_{r}$.
For $l_{r} = 17.5$, the mobility remains low up to $D = 0.2$ and then
begins to increase gradually,
which also correlates with the point at which
the value of $C_{l}$ begins to drop.
For $l_{r} = 75$, the mobility increase starts near $D = 0.4$.
For the
three largest
values of $l_{r}$ there is a tendency
for $M$ to be flat or
even decrease slightly with increasing $D$ up to $D = 0.4$.
This occurs because under moderate obstacle dilution,
the activity-induced clusters
can become more coherent or better ordered.
For $l = 0.175$, there is no clustering and the mobility
increases monotonically  with $D$,
while at $l = 17.5$ some clustering appears for small $D$.
In the regime of larger $l_{r}$,
$C_{l}$ remains large even
for $D = 1.0$
since the clustering or motility-induced phase separation occurs
for these run length values
even in the absence of obstacles.

\begin{figure}
\includegraphics[width=\columnwidth]{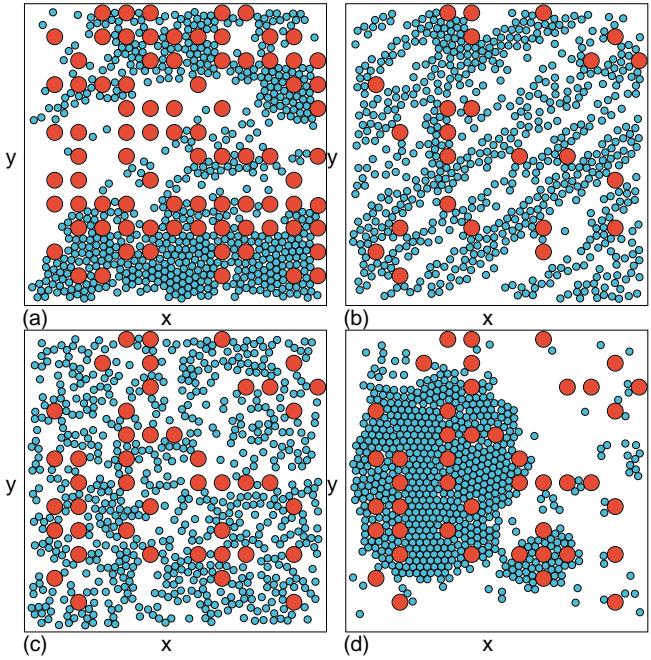}
\caption{
  Snapshots showing the active disk positions (blue) and obstacle locations
  (red) for 
  the diluted system in Fig.~\ref{fig:9} with
  $F_D=0.1$,
  $\phi_{\rm tot}=0.622$, and
  $\theta=35^\circ$.
  (a) A partially clogged state at
  $l_r=0.005$ and $D = 0.41$.
  (b) Reduced clustering for
  $l_r=0.005$ and $D = 0.833$.
  (c) A uniform fluid at
  $l_r=0.175$ and $D = 0.694$.
  (d)
  A clustered state with reduced mobility at
  $l_r=1750$ and $D = 0.694$.
}
\label{fig:10}
\end{figure}

Based on the results in Fig.~\ref{fig:9}, we
can identify several different phases.
At low $l_{r}$ and low dilution $D$, the
system is in a thermally clogged phase with $M \approx 0$,
while at finite dilution the
system enters a partially clogged state
consisting of a combination of flowing and clogged configurations,
as shown in Fig.~\ref{fig:10}(a) for $l = 0.005$ and 
$D = 0.41$. As a result, even though $M$ is finite,  
$C_{l}$ remains large.
In Fig.~\ref{fig:10}(b) we illustrate
the same system at $D= 0.833$ where
$M$ is high and the clustering is strongly reduced, giving
a low value of $C_{l}$.
Figure~\ref{fig:10}(c) shows the uniform fluid
at $l_{r} = 0.175$ and $D = 0.694$,
while in Fig.~\ref{fig:10}(d) at
$l_r=1750$ for the same dilution,
the system forms a clustered state
with a reduced mobility of $M=0.35$.

\begin{figure}
\includegraphics[width=\columnwidth]{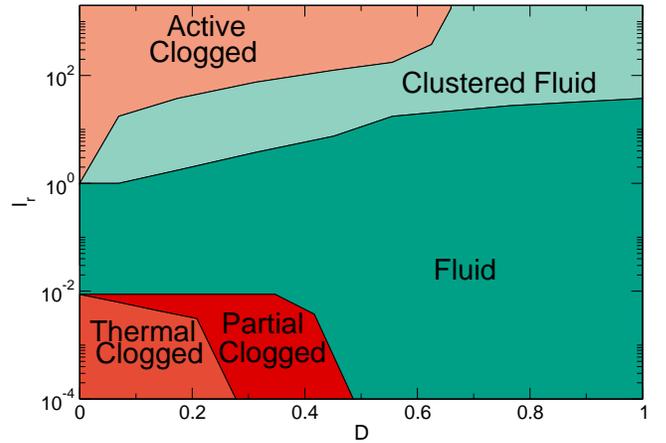}
\caption{ 
Phase diagram as a function of $l_r$ vs dilution fraction $D$
for the system in Figs.~\ref{fig:9} and \ref{fig:10}
with
$F_D=0.1$,
$\phi_{\rm tot}=0.622$, and
$\theta=35^\circ$.
Light red:
a thermally clogged state with $M \approx 0$ but high $C_{l}$.
Dark red:
a partially clogged state
with intermediate $M$ and high $C_{l}$.
Dark green: a fluid phase with high $M$ and low $C_{l}$.
Light green: a clustered fluid with high $M$ and high $C_{l}$.
Orange: the active clogged state with low $M$ and high $C_{l}$.   
}
\label{fig:11}
\end{figure}

From the results in Fig.~\ref{fig:10},
we can construct a phase diagram as a function of $l_r$ versus $D$,
shown in Fig.~\ref{fig:11}.
We identify a 
thermally clogged state in which
$M \approx 0$
and $C_{l}$ is high.
This is adjacent to a partially clogged state of the type 
shown in Fig.~\ref{fig:10}(a)
containing a combination of clogged disks and mobile disks, 
where $M$ is finite but $C_{l}$ is high.
There is a uniform fluid phase
of the type illustrated in
Fig.~\ref{fig:10}(b,c) where the mobility is high and clustering is low,
an active clogged phase
where $M$ is low and $C_{l}$ is high,
and a clustered fluid where $M$ is high and $C_{l}$ is high.  
The phase digram in Fig.~\ref{fig:11}
has several similarities to the phase diagram found in
experiments with colloids moving over random obstacles \cite{Stoop18}.
In particular, as
the 
obstacle density is decreased, both systems are 
in a clogged phase with close to zero motion
when $D$ is low,
transition into partially clogged phases
with intermediate velocities
for intermediate $D$, and reach flowing phases for high $D$.  

\section{Varied Driving Angles}

\begin{figure}
\includegraphics[width=\columnwidth]{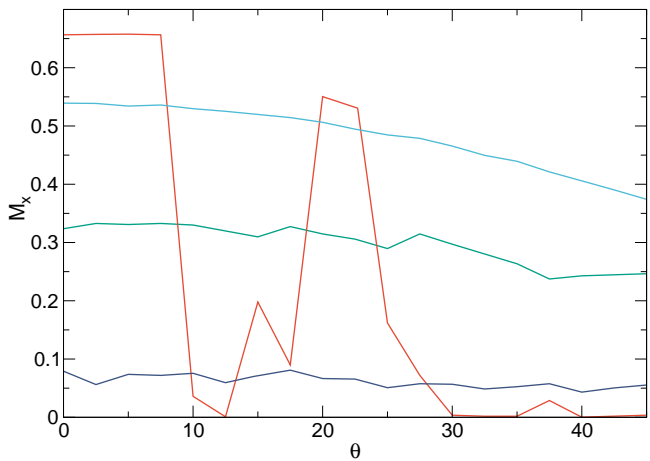}
  \caption{The mobility in the $x$-direction
    $M_x$ vs $\theta$ for the system in
    Fig.~\ref{fig:3} with $F_D=0.1$
    and $\phi_{\rm tot}=0.622$
    at $l_{r} = 0.0025$ (red),
    $0.175$ (light blue), $17.5$ (green),
and $175$ (dark blue) 
}
\label{fig:12}
\end{figure}

In Fig.~\ref{fig:12} we plot the $x$ direction mobility
$M_x=\langle V_x\rangle/F_D$ versus
driving direction $\theta$ for
the system in Fig.~\ref{fig:3} with
$F_D=0.1$
and $\phi_{\rm tot}=0.622$
for varied
$l_{r} = 0.0025$
$0.175$,
$17.5$,
and $175$.
For $l_{r} = 0.0025$,
the mobility
is high for $\theta < 10^\circ$
since the
active disks are moving along
quasi-1D channels similar to those shown in
Fig.~\ref{fig:2}(a).
For $ 10^\circ \leq \theta < 20^\circ$,
a partially clogged state forms,
while for
$20^\circ \leq \theta < 25^\circ$, the active disks flow in
another quasi-1D state.
For  $\theta < 30^\circ$, the
thermally clogged state appears
as shown earlier.
The emergence of
clogged states for certain driving directions
and flowing states for other driving directions
was also observed
in $T = 0$ passive disk studies
\cite{Reichhardt21a}.
For $l_{r} = 0.175$ in Fig.~\ref{fig:12}, the system
forms a uniform fluid for all values of
$\theta$
and there are no clogging phases or locking steps. 
For $l_{r} = 17.5$, some clustering occurs but there is still
no clogging phase, and
the mobility $M_x$ is reduced
at all values of $\theta$.  
At $l_{r}=175$, $M_x$ is even more strongly reduced
for all values of $\theta$ but there
is still no directional locking.
For lower disk densities than those considered in the present work,
steps in the mobility occur
at certain driving angles which correspond
to symmetry directions of the obstacle array \cite{Reichhardt20}.
At the higher disk densities
of Fig.~\ref{fig:12},
the steps disappear.
These results differ from the previously studied collective behavior of
$T = 0$ non-active systems \cite{Reichhardt20a} and from
individual active matter particles
moving on periodic obstacle arrays \cite{Reichhardt20},
where locking steps appeared.
It may be possible that additional
locking states could emerge for lower active disk densities,
smaller obstacle radii, or larger $a$; however, the general trend is that
collective effects reduce directional locking.

\section{Drive Dependence}

\begin{figure}
\includegraphics[width=\columnwidth]{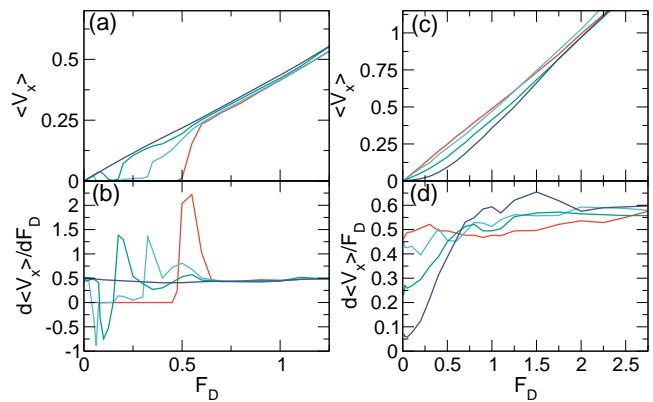}
\caption{ 
  (a) $\langle V_{x}\rangle$ and (b)
  $d\langle V_{x}\rangle/dF_{D}$ vs $F_{D}$
  for the system in Fig.~\ref{fig:3}
  with
  $\phi_{\rm tot} = 0.67$ and $\theta=35^\circ$
  for $l_{r} = 0.00025$ (red),
  $0.0075$ (light blue),
  $0.0175$ (green)
and $0.05$.(dark blue).
(c) $\langle V_{x}\rangle$ and (d) $d\langle V_{x}\rangle/dF_{D}$
vs $F_{D}$ for the same system at
$l_{r} = 0.175$ (red), $1.75$ (light blue), $17.5$ (green),
and $175$ (dark blue).
}
\label{fig:13}
\end{figure}

We next examine the effect on the different phases of
varying the magnitude of the external driving amplitude $F_{D}$.
In Fig.~\ref{fig:13}(a,b) we plot
$\langle V_x\rangle$ and $d\langle V_{x}\rangle/dF_{D}$
versus $F_{D}$ for the system in Fig.~\ref{fig:3} at 
$\phi_{\rm tot} = 0.67$ for driving
along $\theta = 35^\circ$ at
$l_{r} = 0.00025$,
$0.0075$,
$0.0175$,
and $0.05$.
At $l_{r}  = 0.00025$, there is a well defined depinning threshold
near $F_{D} = 0.5$ corresponding to the
break up of the clogged state, which also
appears
as a peak in the corresponding
$d\langle V_x\rangle/dF_{D}$ curve.
For $l_{r} = 0.0075$,
an increase in $\langle V_x\rangle$ from zero occurs near
$F_{D} = 0.15$, followed by
a sharper increase in $\langle V_x\rangle$
at $F_{D} = 0.3$ and
a saturation in the response for $F_{D} > 0.6$. 
The depinning near $F_{D} = 0.15$ results from the
partial breakup of the clogged state into a phase where clogging and flowing
disks coexist,
while at the second stronger
increase in $\langle V_x\rangle$, nearly all of the disks are flowing.

For $l_{r} = 0.0075$ and $0.0175$ in Fig.~\ref{fig:13}(a,b),
the disks immediately begin to flow at small but nonzero $F_D$,
but at slightly higher drives, $\langle V_x\rangle$ drops
nearly back down to zero.
Here, there is a uniform fluid phase
for low $F_{D}$ and finite run lengths
since the drive is not 
strong enough to cause the disks to get stuck behind obstacles.
When $F_{D}$ increases, however,
the driving force overwhelms the activity and
the disks begin
to accumulate behind the obstacles, forming a clogged state in which
the disks are unable to hop backward in order to move around the
obstacles.
The transition from the fluid state to the clogged state produces
negative differential mobility with
$d\langle V_x\rangle/dF_{D} < 0$, while in the 
clogged phase, $d\langle V_{x}\rangle/dF_{D} \approx 0$.
Negative differential mobility has been observed previously for
the single particle limit of passive particles
driven through
random obstacles,
where
sufficiently strong
drives hold
the particle trapped behind an obstacle,
while under
smaller drives the particle can thermally escape from behind the
obstacle
\cite{Leitmann13,Baerts13,Benichou14}. 
In our active system, the
clogged phase
at small but finite drives
extends down to lower drives
as $l_r$ decreases.
Even for $l_{r} = 0.00025$ there is
a flowing fluid phase at drives lower than
those shown in Fig.~\ref{fig:13},
but at $l_{r} = 0$ the initial fluid phase is lost.
For $l_{r} = 0.05$, both the clogging and depinning threshold behavior are lost.

Negative differential mobility
was also observed
for active particles on random obstacle arrays \cite{Reichhardt18},
but there are some differences in the response
compared to what we find for the periodic obstacle arrays.
The negative differential mobility decreases when
the total density of the system is increased
in the random arrays,
as shown in Ref.~\cite{Reichhardt18}, whereas
it increases with increasing total density
in the periodic obstacle
array.
This is because
collective clogging
is responsible for the
negative differential mobility 
in the periodic array,
while individual particle trapping
produces the negative differential mobility in the
random array and in
most other systems which display this effect.

In Fig.~\ref{fig:13}(c,d) we plot
$\langle V_{x}\rangle$ and
$d\langle V_{x}\rangle/dF_{D}$  versus $F_{D}$ for the
system from Fig.~\ref{fig:13}(a,b) at
$l_{r} = 0.175$,
$1.75$,
$17.5$,
and $175$.
For $l_{r} = 0.175$,
the velocity-force curve is almost linear, which is the expected behavior
for the depinning of a fluid \cite{Fisher98}.
Some deviation from linearity appears
at $l_{r} = 1.75$ 
and
becomes
more pronounced at $l_{r} = 17.5$ and $l_r=175$.
The $d\langle V_{x}\rangle/dF_{D}$
curves also show increasing nonlinearity as $l_r$ increases,
with a drop in value at low $F_D$ followed by a linear increase and a
saturation at higher $F_D$.
There is also an overall drop in the velocity
with increasing $l_r$.
The nonlinear behavior in the velocity-force curves
is similar to what is found
for plastic depinning of passive
particle assemblies driven over quenched disorder, such as
colloids or superconducting vortices
\cite{Fisher98,Reichhardt17,Reichhardt02,Fily10}. 
As the run length increases
in the active disk system,
the disks spend
more time in a strongly clustered state and
begin to act
like a solid or glass coupled to 
random disorder.
Increasing the run length is similar to decreasing the temperature
in a glass.
In the limit of
infinite run length,
we expect that there would be a true finite depinning threshold.

\begin{figure}
\includegraphics[width=\columnwidth]{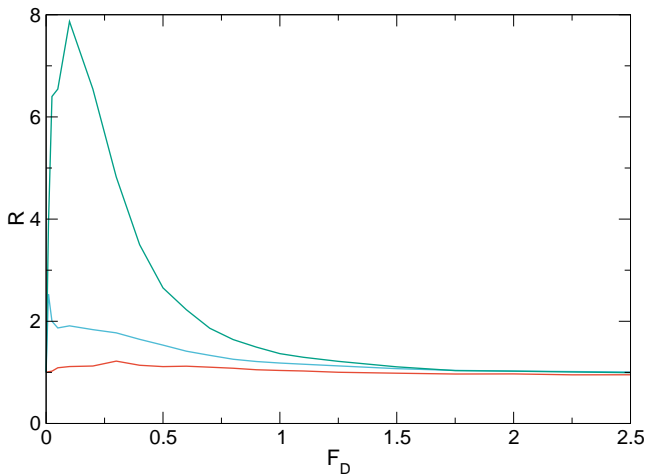}
  \caption{The ratio $R$ of $\langle V_{x}\rangle$ divided by
    $\langle V_x\rangle$ for the $l_r=0.175$ system from Fig.~\ref{fig:13}(c)
    versus $F_{D}$ in samples with
    $\phi_{\rm tot}=0.67$ and
    $\theta=35^\circ$ for
    $l_{r} = 175$ (green),
    $17.5$ (blue),
    and $1.75$ (red),
    showing that $R$ goes to $R=1.0$ at large drives.  
}
\label{fig:14}
\end{figure}

Figure~\ref{fig:13}(c)
shows
that
the differences between the velocity-force curves
for the different activity levels begin to disappear at higher drives.
To make this more clear,
in Fig.~\ref{fig:14} we plot the ratio $R$ of
$\langle V_{x}\rangle$ divided by $\langle V_{x}\rangle$ for the
$l_r=0.175$ sample from Fig.~\ref{fig:13}(c) versus
$F_{D}$
for $l_{r} = 175$,
$17.5$,
and $1.75$.
For $l_{r} =175$, the mobility at low $F_D$ is nearly eight times as large
as in the $l_r=0.175$ system, but
as $F_{D}$ increases, the ratio
of the two mobilities approaches $R=1.0$.
At higher drives,
the collisions with obstacles become strong enough to break apart the
clusters,
while the driving force prevents the active disks
from reorganizing into a cluster state; therefore, the system remains in
a fluid state.
This can also be viewed as a drive dependent
shear thinning transition
in which
the drive breaks up the large scale structures and thereby
reduces the viscosity of the system.

\subsection{Density Dependence of the Driven Dynamics}

\begin{figure}
\includegraphics[width=\columnwidth]{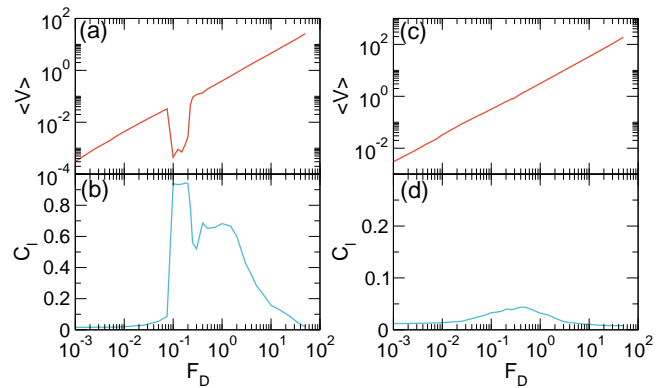}
\caption{ 
  (a) $\langle V_x\rangle$ and
  (b) $C_{l}$ vs $F_{D}$ for a system with
  $\phi_{\rm tot} = 0.622$ and $l_{r} = 0.0075$.
  Here
  $\theta=35^\circ$.
  There are four distinct phases:
  the low drive fluid phase illustrated in Fig.~\ref{fig:16}(a),
  the intermediate drive thermally clogged phase
  shown in Fig.~\ref{fig:16}(b),
  the flowing partially clustered phase appearing in Fig.~\ref{fig:16}(c), and
  the high drive moving liquid phase shown in Fig.~\ref{fig:16}(d).
  (c) $\langle V_x\rangle$ and
  (d) $C_l$ vs $F_D$ for the same system at
  $\phi_{\rm tot} = 0.5238$.
  There are only two phases:
  the liquid phase shown in Fig.~\ref{fig:17}(a) and
  the directionally locked phase flowing along
  $45^\circ$, shown in Fig.~\ref{fig:17}(b).
}
\label{fig:15}
\end{figure}

As $\phi_{\rm tot}$ is reduced,
the extent of the clogged state 
appearing in Fig.~\ref{fig:13}(a,b) is also reduced.
To better characterize both the
clogged phase and the reentrant fluid,
in Fig.~\ref{fig:15}(a,b) we plot $\langle V_{x}\rangle$
and $C_{l}$ versus $F_{D}$
for a system with $\phi_{\rm tot} = 0.622$ and $l_{r} = 0.0075$.
In Fig.~\ref{fig:15}(a),
$\langle V_{x}\rangle$ increases
linearly with increasing $F_D$
up to $F_D=0.08$,
as expected for a fluid, and then
decreases with increasing drive up to $F_{D} = 0.225$.
For $F_{D} < 0.08$
in Fig.~\ref{fig:15}(b),
$C_{l} < 0.03$
and the system forms
a diffusing liquid state without any clustering.
When $0.08 < F_{D} < 0.225$,
$C_{l}$ jumps up to $C_l=0.94$ and the
disks form a system-spanning thermally clogged state.
The velocity does not drop
completely to zero in this clogged phase since there
are still occasional thermal-like jumps of the disks.
For $F_{D} > 0.225$, $\langle V_x\rangle$
increases with increasing $F_D$ again and there
is a drop down in $C_{l}$;
however, there is still a partial clustering effect.
The clustering diminishes as $F_{D}$
increases, and at high drives
$C_l=0.05$
when the system is in a
uniform fluid state.

We define the different phases in Fig.~\ref{fig:15}(a,b)
as a low drive uniform diffusing fluid, a
thermally clogged state,
a partially clustered fluid,
and a high drive fluid state.
The transitions
between the different states appear
as changes in $d\langle V\rangle/dF_{D}$ and $C_{l}$. 
In studies of driven vortices, colloids,
and other particle assemblies \cite{Reichhardt17},
different dynamical phases
also appear as a function of drive; however, in most of these systems
the substrate is
modeled as pinning sites rather than
obstacles,
so
the low drive fluid phase is not present.
Additionally, since
many of these systems have longer range particle-particle interactions,
clustered or clogged states do not occur
since the pairwise energy cost
of strong density inhomogeneities would be too large;
however, the disorder or roughening of the system
is still maximized just at the depinning transition
\cite{Reichhardt17}.
At higher drives, systems with pinning generally show a transition
to an ordered state since the effectiveness of the pinning diminishes
as the driving increases;
however, in systems such as ours containing obstacles,
the overall density becomes
more uniform at higher drives
but the particle positions remain disordered.

\begin{figure}
\includegraphics[width=\columnwidth]{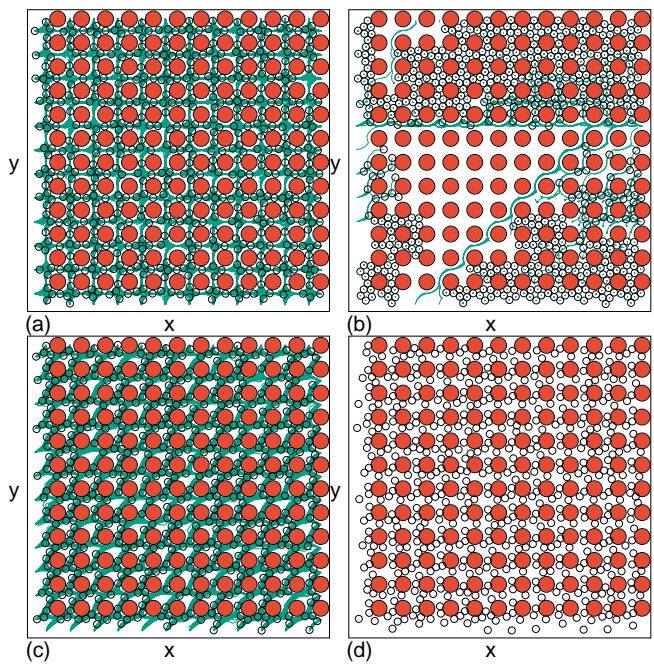}
\caption{The active disk positions (open circles) and trajectories (green) along
  with the obstacle locations (red)
  for the system in Fig.~\ref{fig:15}(a,b) with
  $\theta=35^\circ$, $\phi_{\rm tot}=0.622$,
  and $l_r=0.0075$.
  (a) The moving diffusing fluid phase at $F_{D} = 0.025$.
  (b) The thermally clogged state at $F_{D} = 0.2$.
  (c) The moving partially clustered fluid at $F_{D} = 0.7$.
  (d) The positions of the disks without trajectories in
  the high drive uniform fluid at $F_{D} = 15.0$.
}
\label{fig:16}
\end{figure}

In Fig.~\ref{fig:16}(a) we plot the disk trajectories, obstacle locations,
and active disk positions
for the system in Fig.~\ref{fig:15}(a) at $F_{D} = 0.025$
where the particles form
a low drive diffusing fluid.
Here there is no clustering
and the trajectories are strongly disordered
as the disks gradually drift along $35^\circ$.
Figure~\ref{fig:16}(b) shows the same
system at $F_{D} = 0.2$ in a thermally clogged phase,
where there are still some small regions of
motion.
At $F_D=0.7$ in
Fig.~\ref{fig:16}(c),
most of the disks are flowing in a fluid state
and there is intermittent clustering.
We show only the disk positions without trajectories
in Fig.~\ref{fig:16}(d)
for the high drive uniform fluid at $F_{D} = 15.0$,
where there is no clustering but the disk positions are disordered.

\begin{figure}
\includegraphics[width=\columnwidth]{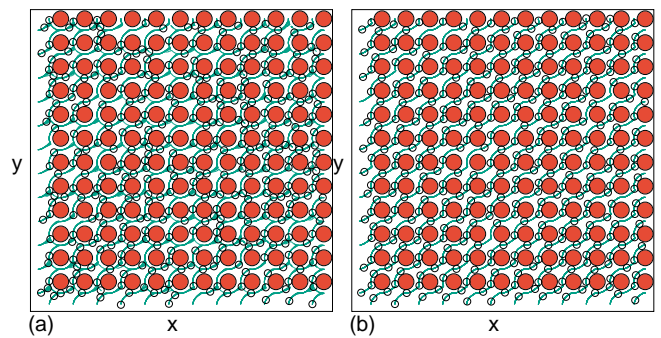}
\caption{The active disk positions (open circles) and trajectories (green)
  along with the obstacle locations (red)
  for the system in Fig.~\ref{fig:15}(c,d)
  with
  $\theta=35^\circ$, $\phi_{\rm tot}=0.5238$,
  and $l_r=0.0075$.
  (a) At $F_{D} = 0.5$, there is a moving liquid with an average direction of
  motion along the driving direction of $\theta = 35^\circ$.
  (b) At $F_{D} = 10.0$, the motion occurs in non-overlapping 1D channels
  locked to $45^{\circ}$.
}
\label{fig:17}
\end{figure}

\begin{figure}
\includegraphics[width=\columnwidth]{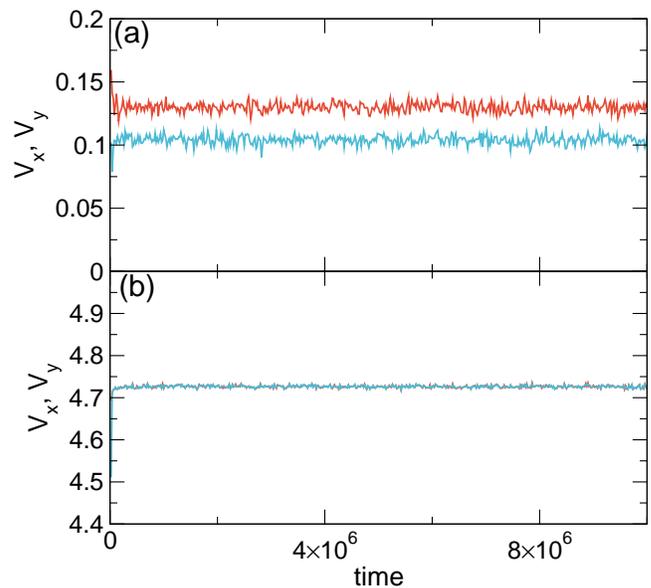}
\caption{The time series of the velocities
  $V_x$ (red) and $V_y$ (blue) for the system in
  Fig.~\ref{fig:15}(c,d)
  with
  $\theta=35^\circ$, $\phi_{\rm tot}=0.5238$,
  and $l_r=0.0075$.
  (a) $F_{D} = 0.5$, where the system forms a
  moving fluid
  translating along the driving direction
  of $\theta=35^\circ$, as shown in Fig.~\ref{fig:17}(a).
  (b) $F_{D} = 10$, where the motion is locked along
  $45^\circ$
  and occurs in 1D non-overlapping channels.
}
\label{fig:18}
\end{figure}

In Fig.~\ref{fig:15}(c,d) we plot
$\langle V_{x}\rangle$ and $C_{l}$
for the system in Fig.~\ref{fig:15}(a,b)
at a lower $\phi_{\rm tot} = 0.5238$. In this case
there is no clogged phase or clustering, indicating
that there is a critical value of $\phi_{\rm tot}$
below which the clogging phase
cannot occur.
For low values of $\phi_{\rm tot}$ at lower $F_{D}$, 
the system forms a moving liquid phase in which
the average motion is along
the driving direction of $\theta=35^\circ$
but individual disks can diffuse over time in all directions,
as shown in Fig.~\ref{fig:17}(a) for $F_{D} = 0.5$.
In Fig.~\ref{fig:18} we plot
the time series of $V_{x}$ and $V_{y}$ for the system in
Fig.~\ref{fig:17}(a). The velocities are fluctuating but the velocity
components do not overlap.
At higher drives, there is a transition
from the diffusing fluid into a directionally locked phase
with motion along $45^\circ$ that is strictly
quasi-1D without hopping between
channels, as shown in
Fig.~\ref{fig:17}(b) at $F_{D} = 10.0$.
Here $V_x$ and $V_y$ have identical average values,
as shown in Fig.~\ref{fig:18}(b).
At even lower densities or
for smaller obstacle sizes, additional directional locking phases appear.

\begin{figure}
\includegraphics[width=\columnwidth]{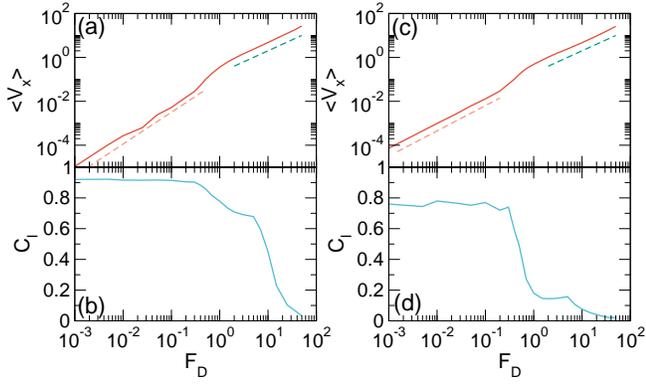}
\caption{ 
(a) $\langle V_{x}\rangle$ and (b)
  $C_{l}$ vs $F_{D}$ for a system with
  $\theta=35^\circ$,
  $\phi_{\rm tot} = 0.622$ and $l_{r} = 175.0$. 
  The orange dashed line is a fit to
  $\langle V_{x}\rangle \propto F^{1.45}$,
  and the green dashed line indicates a linear
  dependence of $\langle V_x\rangle$ on $F_{D}$.
  (c) $\langle V_x\rangle$ and (b) $C_l$ vs $F_D$ for the same
  system
  at $\phi_{\rm tot} = 0.523$.
  The orange dashed line is a fit to
  $\langle V_x\rangle \propto F^{1.15}$ and the green dashed line is a linear
fit.  
}
\label{fig:19}
\end{figure}

In Fig.~\ref{fig:19}(a,b) we plot
$\langle V_{x}\rangle$ and $C_{l}$ versus $F_{D}$
for the system in Fig.~\ref{fig:15}(a,b) with 
$\phi_{\rm tot} = 0.622$ at a higher $l_{r} = 175$.
Since the clustering in this regime is due to the activity,
$C_{l}>0.9$ down to arbitrarily low values of $F_{D}$. This is
in contrast to
the drive dependent clogging effect
which appears at smaller $l_{r}$.
Near $F_{D} = 0.5$, there is a change in the velocity-force curve
corresponding to a drop
in $C_{l}$, while
a second small
cusp in $C_{l}$ is visible near $F_{D} = 7.0$.
For $F_{D} < 0.5$, the system is in the active clogged state
and the velocity-force curve
is nonlinear, as indicated by the orange dashed line which
shows a power law fit to
$\langle V_x\rangle \propto F_{D}^{1.45}$.
Above the
cusp in
$\langle V_{x}\rangle$, the green dashed line indicates that the behavior
is linear with
$\langle V_x\rangle \propto F_{D}$.
For passive 2D systems with quenched disorder at $T=0$,
plastic flow regimes
are associated with a velocity-force signature of
$V \propto F^\alpha$ with $1.33 < \alpha < 2.0$,
where $\alpha \approx 1.5$ is often observed
\cite{Fisher98,Reichhardt17,Fily10}.
This suggests that the active clogged regime behaves
like a plastically distorting solid moving over quenched disorder.
At higher drives, the self clustering starts to break down
and the velocity response becomes linear,
as expected for a strongly driven fluid
in quenched disorder
\cite{Benichou14,Reichhardt17}.

In Fig.~\ref{fig:19}(c,d) we plot $\langle V_x\rangle$ and $C_l$
versus $F_D$
for the system from Fig.~\ref{fig:19}(a,b) at
a lower $\phi_{\rm tot} = 0.523$.
We find a similar behavior in which
the system starts out in a clustered state at low drives
and transitions to a moving fluid at higher drives.
There is a cusp 
in $\langle V_{x}\rangle$ near $F_{D} = 1.0$
which corresponds to a drop in $C_{l}$.
Below the cusp, $\langle V_x\rangle \propto F^{1.15}$,
while above the cusp,
$\langle V_x\rangle \propto F^{1.0}$. This suggests that
although clustering is occurring at the lower drives,
the cluster sizes are smaller compared to the $l_r=175$ case.
As a result,
the system behaves more like a fluid
than like a plastically moving solid. 
There is also a small cusp in $C_{l}$ near $F_{D} = 5.0$
which corresponds to the transition
from a moving fluid to locked flow along
$45^\circ$.

\begin{figure}
\includegraphics[width=\columnwidth]{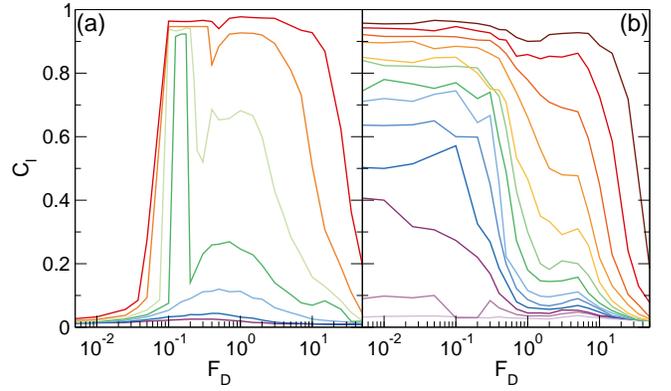}
\caption{ 
  (a) $C_{l}$ vs $F_{D}$ for a system with
  $\theta=35^\circ$ at
  a fixed $l_{r} = 0.0075$ for
  varied $\phi_{\rm tot} = 0.671$ (red),
  0.6465 (orange),
  0.622 (light green),
  0.59744 (dark green),
  0.5729 (light blue),
  0.5238 (dark blue), and
  $0.475$ (purple), from top to bottom.
(b)
$C_{l}$ vs $F_{D}$ for the same system at $l_{r} = 175$ at
  $\phi_{\rm tot} = 0.671$ (brown),
  0.6465 (red),
  0.622 (dark orange),
  0.59744 (medium orange),
  0.5729 (light orange),
  0.548 (light green),
  0.5238 (dark green),
  0.5 (light blue),
  0.475 (medium blue),
  0.45 (dark blue),
  0.4256 (dark purple),
  0.401 (medium purple), and
  $0.3765$ (light purple), from top to bottom.
}
\label{fig:20}
\end{figure}

In Fig.~\ref{fig:20}(a) we plot $C_{l}$ versus $F_{D}$
for a system with fixed $l_{r} = 0.0075$ at
varied $\phi_{\rm tot} = 0.671$, 0.6465, 0.622, 0.59744, 0.5729, 0.5238,
and $0.475$.
The four phases are clearly visible.
In the low drive fluid phase, $C_{l} < 0.1$. The clogged phase corresponds
to the peak in $C_{l}$,
while the partially clustered liquid with $C_{l} > 0.5$ appears
above
the clogging peak.
At high drives, there is a liquid
with $C_{l} < 0.5$.
The width of the clogged phase decreases with decreasing
$\phi_{\rm tot}$.
For $\phi_{\rm tot} = 0.59744$, the system passes
directly from the clogged state
to the liquid phase, while 
for $\phi_{\rm tot} < 0.59744$, the
system is always in a fluid state at lower drives and reaches
a directionally locked phase with motion along $45^\circ$ at higher drives.

\begin{figure}
\includegraphics[width=\columnwidth]{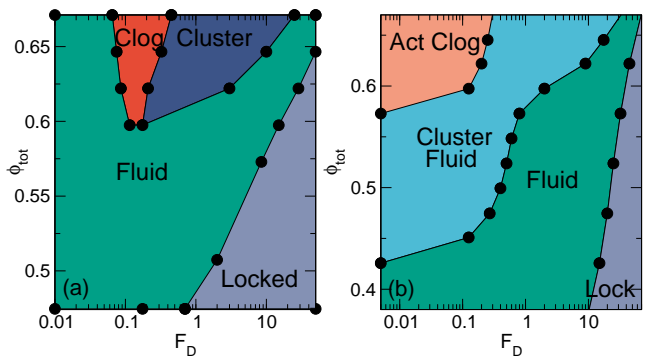}
\caption{ 
  (a) Phase diagram as a function of $\phi_{\rm tot}$ vs $F_D$
  for the system in Fig.~\ref{fig:20}(a)
  with
  $\theta=35^\circ$
  at $l_{r} = 0.0075$, showing the
  reentrant fluid phase (green),
  the thermally clogged phase (red),
  the partially clustered moving phase (blue) and the
directionally locked phase (light purple).   
(b) Phase diagram as a function of $\phi_{\rm tot}$ vs $F_D$
for the system in Fig.~\ref{fig:20}(b) with $l_r=175$,
showing the active clogged phase (orange),
the clustered fluid (blue), the fluid phase (green), and 
the directionally locked phase (light purple).
}
\label{fig:21}
\end{figure}

In Fig.~\ref{fig:21}(a) we plot
a phase diagram as a function of $\phi_{\rm tot}$ versus
$F_{D}$ for the system in Fig.~\ref{fig:20}(a) with $l_r=0.0075$.
The transitions between the phases are obtained based on
the behavior of
$C_{l}$, the velocities, and the trajectories.
At low $F_{D}$ and low $\phi_{\rm tot}$ we find a fluid phase, while at high
$F_{D}$ a 1D locked phase with motion along 45$^\circ$ emerges. 
For $\phi_{\rm tot} \geq 0.59744$,
both clogged and reentrant fluid phases can occur.
The boundaries of the clustered fluid phase shift
at different rates to higher $F_D$ with
increasing $\phi_{\rm tot}$, causing the clustered fluid
to increase in extent for larger $\phi_{\rm tot}$.
The transition from the fluid to the symmetry
locked flow along 45$^\circ$ shifts to
lower drives with decreasing $\phi_{\rm tot}$.  

In Fig.~\ref{fig:20}(b) we plot $C_{l}$ versus $F_{D}$ 
for the system in Fig.~\ref{fig:20}(a) at $l_{r} = 175$
for 
$\phi = 0.671$, 0.6465, 0.622, 0.59744, 0.5729, 0.548,
0.5238, 0.5, 0.475, 0.45, 0.4256, 0.401, and $0.3765$.
We find clustering with $C_{l} > 0.5$ for lower drives
when $\phi_{\rm tot} > 0.4256$.
The transition from the cluster
phase to a moving state shifts to lower $F_{D}$ as
$\phi_{\rm tot}$ is reduced.
Additionally, for higher drives
the system enters the 1D locking
phase with motion along $45^\circ$.
In general, as $l_{r}$ increases, a higher driving
force is needed to reach the locked state.
In Fig.~\ref{fig:21}(b) we plot a phase diagram as a function of
$\phi_{\rm tot}$ versus $F_D$
for the system in Fig.~\ref{fig:20}(b) highlighting
the four phases.
The active clogged phase
is defined to occur when $M$ is below 15\% of its maximum
value and $C_{l}$ is large. 
The clustered phase appears when $C_{l} > 0.5$ and
$M$ is greater than 15\% of its maximum value.
In the fluid phase, the mobility is high and there is no
clustering, while in the locked phase, there
is 1D motion along $45^\circ$.
Compared to the system with a lower value of $l_r$ in
Fig.~\ref{fig:21}(a), when $l_r$ is higher,
the clogging and cluster phases
extend to lower $\phi_{\rm tot}$ and larger $F_{D}$.

\begin{figure}
\includegraphics[width=\columnwidth]{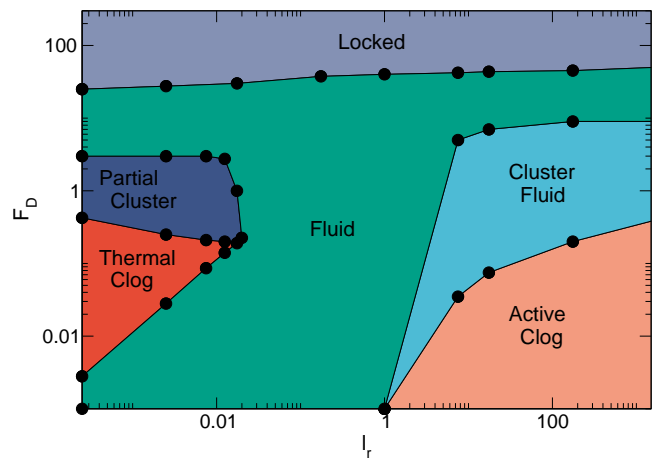}
\caption{ 
The phase diagram as a function of
$F_{D}$ vs $l_{r}$ for the system in Fig.~\ref{fig:20} with
$\theta=35^\circ$
for a fixed $\phi_{\rm tot} = 0.622$,
showing the thermally clogged phase (red),
the fluid phase (green),
a partially clustered fluid phase (dark blue),
an active clogged phase (orange),
an active clustered fluid phase (light blue),
and the directionally locked 1D flow
along $45^\circ$ (light purple). 
}
\label{fig:22}
\end{figure}

In Fig.~\ref{fig:22} we plot a phase digram as a function of
$F_{D}$ versus $l_{r}$ for the system in Fig.~\ref{fig:21} at
fixed $\phi_{\rm tot} = 0.622$
showing the relationships between 
the thermally clogged phase, the 
reentrant fluid phase,
and the active clogged phase.
At finite but low $F_{D}$ and small $l_{r}$,
the thermally clogged phase
appears,
and its onset
shifts to
higher $F_D$
with increasing
$l_{r}$ 
until
it is completely lost
for $l_{r} > 0.025$.
The clogged phase extends all the way down to $F_D=0$ only when
$l_{r}$ reaches zero.
Under increasing $F_D$,
there is a crossover from
the zero mobility thermally clogged phase
to the low mobility
partially clogged phase
composed of coexisting clogged and moving disks.
Another transition occurs at higher drives
to a uniform fluid with high $M$ and low $C_{l}$. 
When $l_{r} > 1.0$,
an active clogged state and active clustered
fluid appear for low drives.
Unlike the thermally clogged state,
the active clogged state can persist down to $F_{D} = 0$ since 
the clustering is produced by the activity
instead of by the driving.
For all values of $l_r$,
once $F_D$ becomes sufficiently large,
the system enters a locked phase of
1D flow along $45^\circ$.
For driving along $\theta = 0^\circ$,
the thermally clogged and partially
clogged phases at small $l_r$ are
absent, while
the locked flow along $45^\circ$ is
replaced by locked flow
along $0^\circ$.
For other driving angles $\theta$,
the widths of the thermally clogged,
partially clogged, and fluid phases vary,
but the active clogged phases persist.

\section{Summary} 
We have numerically examined
run-and-tumble active disks moving over a periodic obstacle array
in the collective limit. For driving along the
$x$ axis, we find that the system forms
an active clustered state and an active clogged state at
large run lengths.
The onset of clustering
is associated with a drop in the mobility.
For driving at an incommensurate angle,
we find two distinct clogging phases:
a heterogeneous thermally clogged state at low activity,
and an active clogged state at large activity.
The two phases are separated by a uniform fluid state, and
there is an
active clustered state between the active clogged state and the
fluid state.
The thermally clogged state is strongly sensitive to whether the driving
is along a symmetry direction of the substrate or whether it
is along an incommensurate angle,
while the active clogged state is independent of
the driving direction.
The thermally clogged state 
is always system-spanning and involves the formation of a
single large cluster,
while the active clogged state contains multiple smaller clusters
which are scattered throughout the sample.
When the obstacle lattice is diluted via the random removal of
a fraction of the obstacles, 
there is a critical dilution fraction above which
the thermally clogged state is lost; however,
the active clogged state is much more robust against dilution.
We observe a reentrant fluid phase
as a function of driving force in the low activity
regime.
For finite activity,
the system can always flow
when the drive is low enough,
but as the drive increases,
a drive dependent clogged state appears that can
depin for sufficiently high driving.
This drive dependent clogged state is associated with
negative differential mobility in the velocity-force curves.
Above depinning,
there is a partially clogged state consisting of coexisting
clogged and moving regions,
while a moving uniform fluid appears
at higher drives.
In the active clogging regime,
the velocity-force curves are nonlinear and
can be fit to $V \propto F^{\alpha}$ with $\alpha > 1.33$,
similar
to the behavior observed in
plastically flowing passive solids
moving over quenched disorder.
At higher drives, the active clusters break apart
and a moving fluid appears.
We map out phase diagrams to identify the locations of the
reentrant fluid state,
thermal clogging,
a partially clustered phase,
active clogging, and an active clustered state.
We also find that at high drives,
the system undergoes directionally locked
flow through 1D channels aligned with a symmetry
direction of the obstacle lattice.
Our results demonstrate
how a driven active system can transition from thermal clogging behavior
to active clogging behavior as a function of increasing activity.  

\begin{acknowledgments}
We gratefully acknowledge the support of the U.S. Department of
Energy through the LANL/LDRD program for this work.
This work was supported by the US Department of Energy through
the Los Alamos National Laboratory.  Los Alamos National Laboratory is
operated by Triad National Security, LLC, for the National Nuclear Security
Administration of the U. S. Department of Energy (Contract No. 892333218NCA000001).
\end{acknowledgments}

\bibliography{mybib}

\end{document}